\documentclass[a4paper]{jpconf}
\usepackage{graphicx}
\usepackage{amssymb}
\usepackage{cite}

\newcommand{\bea}{\begin{eqnarray}}
\newcommand{\eea}{\end{eqnarray}}
\newcommand{\be}{\begin{equation}}
\newcommand{\ee}{\end{equation}}

\newcommand{\mev}{{\rm MeV}}

\newcommand{\kl}{K_{\ell 3}}
\newcommand{\vus}{\vert V_{us}\vert}

\newcommand{\fp}{f_+(0)}

\def\slash#1{\setbox0=\hbox{$#1$}\dimen0=\wd0                    
      \setbox1=\hbox{/} \dimen1=\wd1 \ifdim\dimen0>\dimen1
      \rlap{\hbox to \dimen0{\hfil/\hfil}} #1                        \else                                       
      \rlap{\hbox to \dimen1{\hfil$#1$\hfil}}  
      /   \fi}                                         
\begin{document}
\title{Precision tests with $K_{\ell3}$ and $K_{\ell2}$ decays}

\author{Federico Mescia~\footnote{{\it On behalf of  the Kaon working group
activity - Flavianet. 
%The members aware of the data used for this write-up are M.~Antonelli, C.~Gatti, G.~Isidori and M.~Moulson.  
The web-page~\cite{flavia} is steadily updated.}}}

\address{INFN, Laboratori Nazionali di Frascati, Via E.
          Fermi 40, I-00044 Frascati, Italy.}
\ead{Federico.Mescia@lnf.infn.it}

\begin{abstract} 

The analysis made in 2000 indicated
that the unitarity relation   $\vert V_{ud}\vert^2 +\vert V_{us}\vert^2 +
\vert V_{ub}\vert^2 = 1$ might be broken at the $2.3\sigma$
level.  At that time, however,  $\vert V_{us}\vert$ was inferred from  old experimental
data.   Since then, a great experimental and theoretical effort has been invested to
understand  the source of that discrepancy. Thanks to the new and improved measurements  by BNL-E865, KLOE,
KTeV, ISTRA+ and NA48,  the old $K_{\ell3}$ decay rate got shifted so that  the  new $\vert
V_{us}\vert$ is now  consistent with unitarity. On the theory side, much progress  in the
lattice QCD  has been made in order to tame  the systematic uncertainties related to
the computation  of the $K_{\ell3}$ form factors.  

This joint progress allowed to assess the validity of the CKM unitarity 
relation at the level of 
less than $1\%$.  The key challenge of the future lattice studies will be to simulate
lighter pions  in the region in which ChPT predictions apply.
Also interesting is the recent progress in accurately computing 
the  kaon and pion decay constants on the lattice, which then give us access to $\vert V_{us}\vert$ and $\vert
V_{ud}\vert$ from the corresponding leptonic decays.

In addition, we discuss that  the  $K_{\ell3}$ and $K_{\ell2}$ decays offer 
the possibility to test various scenarios of physics beyond Standard Model.

\end{abstract}

\section{Introduction}

From the experimental information on down- to up- quark transitions
(such as $d\to u$, $s\to u$ and $b\to u$), we accede the effective 
dimesion-six operators of the form, $D\Gamma_1 U \ell \Gamma_2\nu$, with 
$D$ ($U$) being a generic ``{\it down}" (``{\it up}") flavor, and  $\ell \in \{e,\mu,\tau\}$.  
Their effective coupling $G^2_{UD}$
can be parametrized as the Standard Model (SM) contribution, 
$G_F^2\,\vert V_{UD}\vert^2$, plus a possible new physics 
terms, $G_F^2\,\epsilon_{NP}$. Since  the dimension-six operators are not protected 
by gauge invariance the possible  effects of non-decoupling are proportional to  
$(1+ M^2_W/\Lambda^2_{NP}$).  The effects of these
non-standard contributions cannot be very large, 
but they can become detectable in high-precision
experiments.

A convenient strategy 
to measure these effects against the SM parameters, $G_F^2$ and 
$\vert V_{UD}\vert$,  is to test the Cabibbo universality hypothesis 
(or the unitarity constraint) between quark and lepton:
\begin{equation}
G^2_{CKM}=G^2_\mu\,,\,\quad \left[\mbox{or}\:\:\:
|V_{ud}|^2+|V_{us}|^2+|V_{ub}|^2=1\:,\:\:\mbox{and}\: \:\:G_F\equiv G_\mu \right]\quad ,
\label{eq:unitarity}
\end{equation}
where $G^2_{CKM}=G^2_{ud}+G^2_{us}+G^2_{ub}$, and 
$G_\mu = 1.166371 (6) \times 10^{-5} {\rm GeV}^{-2}$ is $G_F$, as measured from the 
accurate value of the  muon
lifetime~\cite{mulan}. 

We report on the progress related to the verification of the unitarity relation~(\ref{eq:unitarity}), particularly emphasizing 
the progress in taming the underlying hadronic uncertainties. 
As we shall see 
the CKM unitarity relation~(\ref{eq:unitarity}),
is tested at the $1\ \%$ level (and even less), which therefore becomes an important constraint 
for  beyond 
SM physics scenarios.  For example, in SO(10) grand unification theories, the CKM unitarity relation~(\ref{eq:unitarity}) can 
be used to set the bound on the mass of $Z^\prime$, namely  $m_{Z^\prime}$$>$$1.4$~TeV, which is more than competitive with
the one set through the direct collider searches, $m_{Z^\prime}$$>$$720$~GeV~\cite{pdg}.
The unitarity also provides a useful constraint in various supersymmetry breaking scenarios~\cite{barb}.

In what follows, I discuss the present status of $\vert V_{us}\vert$, as obtained from the studies of semileptonic ($K_{\ell3}$)
and leptonic ($K_{\ell2}$) decays. I will mainly concentrate  on the theoretical progress. For the experimental novelties the reader is encouraged to consult the other contributions of the Flavianet Kaon Working
group~\cite{flavia,moulson,wanke,palutan}.  I will also present some prospects for making the new
physics searches from $K_{\ell2}$ decays.

\section{$\vert V_{us}\vert $ and the CKM test: $K_{\ell3}$}

\label{sec:formula}

In the SM, we deal with the following master formulas for $K_{\ell3}$ and 
$K_{\ell2}$ decay rates:
\bea
\label{eq:one}
\Gamma(K_{\ell 3(\gamma)}) &=& 
{ G_\mu^2 M_K^5 \over 128 \pi^3} C_K   
  S_{\rm ew}\,|V_{us}|^2 f_+(0)^2\,
I_K^\ell(\lambda_{+,0})\,\left(1 + \delta^{K}_{SU(2)}+\delta^{K \ell}_{\rm
em}\right)^2\,\quad ,\\
\frac{\Gamma(K^{\pm}_{\ell 2(\gamma)})}{\Gamma(\pi^{\pm}_{\ell 2(\gamma)})} &=& 
\large\left|\frac{V_{us}}{V_{ud}}\large\right|^2\frac{f^2_K m_K}
{f^2_\pi m_\pi}\left(\frac{1-m^2_\ell/m_K^2}{1-m^2_\ell/m_\pi^2}\right)
\times\left(1+\delta_{\rm
em}\right)\quad ,\label{eq:two}
\eea
where  $C_{K}=1$  ($1/2$) for the neutral (charged) kaon decay.
$I_K^\ell(\lambda_{+,0})$ is the phase space integral which also includes the integration of the shape of the form factors 
parametrized by $\lambda_{+,0}$.  The universal short-distance electromagnetic 
correction, $S_{\rm ew}=1.0232(3)$,  has been computed 
 in ref.~\cite{Sirlin:1981ie}, while
 the long-distance electromagnetic corrections 
 $\delta_{\rm em}=0.9930(35)$ and $\delta^{K \ell}_{\rm
em}$, as well as  the  isospin-breaking ones, 
$\delta^{K}_{SU(2)}$, have been  computed in refs.~\cite{marc1,Cirigliano}
(see table~\ref{tab:iso-brk} for $\delta^{K \ell}_{\rm
em}$ and $\delta^{K}_{SU(2)}$). 
%%%%%%%%%%%%%%%%%%%%%%%%%%
%%%%%%%%%%tab:iso-brk
\begin{table}[htb]
\setlength{\tabcolsep}{3.8pt}
\centering
\begin{tabular}{c||c||c|}
& $\delta^K_{SU(2)} (\%)$ 
& $\delta^{K \ell}_{\rm em}(\%) $   \\
\hline
$K^{+}_{e3}$    & 2.36(22)  &  +0.08(15)\\
$K^{0}_{e 3}$   &  0        &  +0.57(15)\\
$K^{+}_{\mu 3}$ & 2.36(22)  &  -0.12(15)          \\
$K^{0}_{\mu 3}$ &  0        &  +0.80(15)\\
\end{tabular}
\caption{Summary of the isospin-breaking
factors~\cite{Cirigliano}}
\label{tab:iso-brk}
\end{table}
The remaining quantities,
 $f_{+}(0)$,  the vector form factor at  zero momentum transfer [$q^2=(p_K-p_\pi)^2 
= 0$], and $f_K/f_\pi$, the ratio of the kaon and pion decay constants, encode the non-perturbative QCD information on the 
flavor SU(3)  breaking effects arising in the relevant hadronic matrix element.

This year, values of all the branching ratios of both neutral and charged $\kl$ decay modes  
from the new kaon experiments became available~\cite{exp}.  When  translated into the 
uncertainty in $\vert V_{us}\vert f_+(0)$, it is only $0.4\%$ for the charged modes 
and $0.1\%$ for the neutral ones~\cite{moulson}.  
Averaged, that uncertainty is about $0.2\%$,  which leads us to 
 $\vert V_{us}\vert f_+(0)=0.21666(48)$. 

\begin{figure}[t]
\centering
\includegraphics[width=0.6\textwidth]{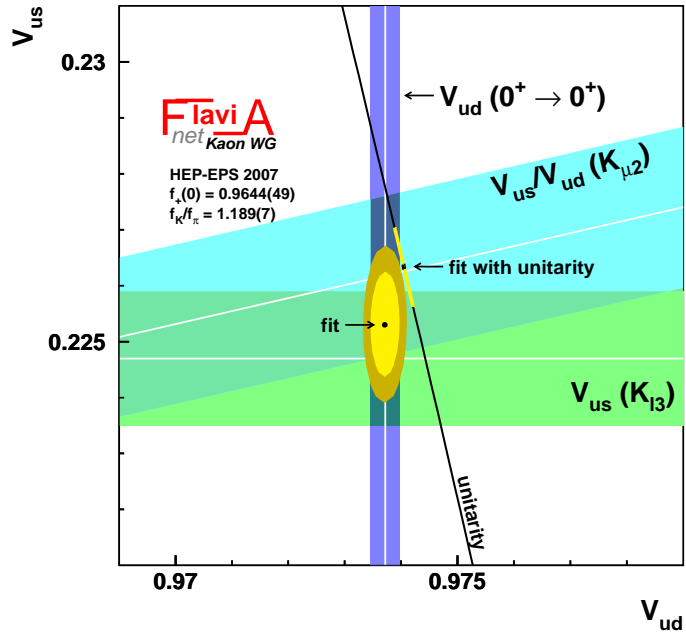}
\vspace*{-1.2cm}
\caption{\label{vusuni} Summary of CKM unitarity test. 
To study the impact of high-precision
lattice data,  we use
$f_+(0)=0.964(5)$~\cite{rbcf0} and $f_K/f_\pi=1.189(7)$~\cite{davies} and we
obtain $V_{ud}^2+V_{us}^2-1=0.0011(7)$. 
 As discussed in the text, all the lattice
groups are currently working to reduce the remaining systematic
uncertainties.}
\end{figure}

For the time being, such a highly precise measurement could not be translated to a similar error on 
 the $\vert V_{us}\vert $ determination. The obstacle is obviously the difficulty to keep 
the theoretical uncertainties in  $f_+(0)$ at the per-mil  level.
In eq.(\ref{eq:one}), $\fp$  is defined in the absence
of electromagnetic corrections and  of the isospin-breaking terms. It is solely due to the strong interactions, described by non-perturbative QCD. 
In the flavor SU(3) limit it is merely equal to unity thanks to the conservation of the vector current. Its deviation 
from that limit is conveniently written as
\vspace*{-0.15cm}
 \be\label{chpt4}
\vspace*{-0.15cm}
 f_+(0)= 1 + f_2 + f_4 + \ldots
 \ee
In chiral perturbation theory (ChPT)  $f_2$ and $f_4$ are the leading and 
next-to-leading chiral corrections respectively.  Ademollo--Gatto theorem ensures that the term $\propto 
(m_s-m_u)$ is absent and thus  $f_2=-0.023$  is an unambiguous prediction 
of ChPT.   
The calculation of the chiral loop contribution, $\Delta(\mu)$ in
\be 
\renewcommand{\arraystretch}{0.5} 
f_4 =
\Delta(\mu) + f_4\vert^{loc}(\mu)\,, 
\label{eq:f4ch}
\ee 
has been recently completed in ref.~\cite{tal}, but the full determination of $f_4$ 
necessitates an accurate estimation of the local counter-term 
 $f_4\vert^{loc}(\mu)$, which is ${\cal O}(p^6)$.
\begin{figure}[t]
\centering
\includegraphics[width=0.7\textwidth]{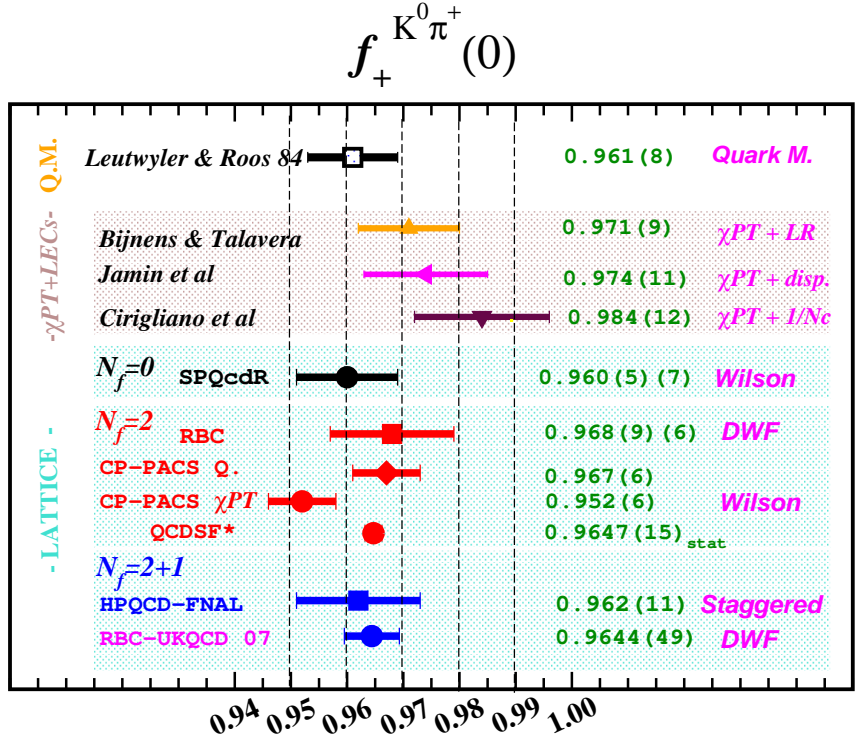}
\caption{\label{f0}
Current situation with the theoretical estimates of $f_+(0)\equiv
f_+^{K^0\pi^-}(0)$~\cite{LR,rbcf0,milcf0,f0}. Each method is being used to
evaluate $f_4$ in eq.~(4), while $f_2=-0.023$ is predicted in
ChPT, in terms of Kaon and pion masses only. 
}\end{figure}
\begin{figure}[t]
\centering
\includegraphics[width=0.8\textwidth]{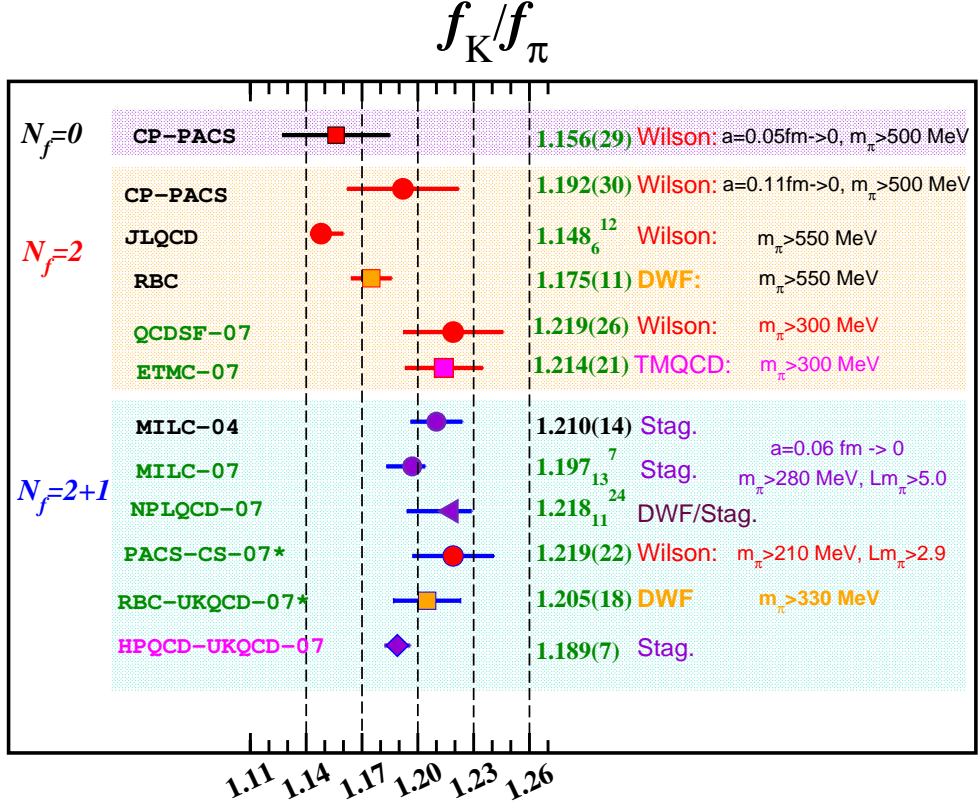}
\caption{\label{fk}Summary of $f_K/f_\pi$
estimates~\cite{milcfk,davies,rbcfk,fk,jam}. All values are from Lattice QCD.               
Recent results relied on the use of ChPT expressions to
extrapolate to the physical limit.
}\end{figure}
Basically, over  the years two theoretical approaches have been used. In one method,
$f_4\vert^{loc}(\mu)$ of eq.~(\ref{eq:f4ch}) is estimated by QCD models such as dispersion relation,
 $1/N_c$ limit and resonance
saturation, whereas in the latter the full $f_4$ is estimated by Lattice QCD.
 All results (see fig.\ref{f0}) essentially confirm the old 
 estimate made  by Leutwyler and Roos which was obtained in 
a simple quark model~\cite{LR}. The benefit of  new results, obtained using 
more sophisticated approaches, lies in the fact that we are nowadays in the position to control the 
systematic uncertainties of our calculations while with the quark models
this is not possible.  To stress the importance of the accurate determination of $f_4$, we 
should remind the reader that the experimental error on $\vert V_{us}\vert f_+(0)$ is only $0.2\%$, whereas the spread of
 theoretical estimates of $\fp$ is still at the $1\% \div2\%$ which is unsatisfactory.
Recent progress in lattice QCD gives us more optimism as far as the prospects of reducing the error on $\fp$ to well 
below $1\%$ are concerned~\cite{kan}.  Most of the currently available results obtained by using  lattice 
QCD  worked with ``heavy pions".  One may notice that the lattice QCD results are  lower
than those obtained by the ChPT-inspired models. 
An important step to improve the accuracy of $\fp$ estimates
has been recently made by 
the UKQCD-RBC collaboration~\cite{rbcf0}. 
Their preliminary  result, $\fp=0.964(5)$
is obtained from the unquenched study  with 
$N_F=2+1$ flavors of quarks which have good chiral properties 
on the lattice (so called, Domain Wall quarks), 
and their pions ($\gtrsim 330\,\mev$) are lighter than those  reported in  previous 
lattice QCD studies. Their 
overall error is estimated to be  $~0.5\%$, which is very encouraging. 
It is important to emphasize that they observe 
a mass dependence similar to that of $f_2$. That is something new 
with respect to previous lattice studies (this is likely
due to the fact that they work with lighter pions). 
One should keep in mind, however, that their result is obtained 
from the simulation at  
a single value of the lattice spacing (i.e. $a=0.12$ fm) and in a relatively small 
extension of the fifth dimension of the lattice
box~\footnote{Even though $m_\pi L\gtrsim 4.5$, 
 simulations with a larger fifth dimesion, $L_s$ would help because  
the mass of lightest quark ($=0.005$ in lattice units)  is very close to the
 residual mass parameter ($=0.003$, also in lattice units). 
 This is, in particular, relevant for $f_K/f_\pi$, which is the order parameter of the
 chiral symmetry.}.
If the RBC-UKQCD esti\-ma\-te, $\fp=0.964(5)$, is combined with the experimental 
ave\-ra\-ge, $ \vus \fp=0.21666(48)$~\cite{flavia}, one gets  that the
 CKM unitarity is confirmed to a precision 
well below  $1\%$ (see fig.~\ref{vusuni}), which is a new result.

A complementary research to provide the accurate estimate of $\vus$ is made through 
the  $K_{\ell2}$ decays. 
The most important mode is  $K^+\to\mu^+\nu$ which has been recently updated by KLOE, so that the 
relative uncertainty is now  $3\%$.  
To minimize the hadronic uncertainties,  in eq.~(\ref{eq:two}) we have introduced the ratio 
 $\Gamma(K^+\to\mu^+\nu)/\Gamma(\pi^+\to\mu^+\nu)$.
In this case, the QCD uncertainty enters with
 \be\label{chpt3}
\vspace*{-0.15cm}
 f_K/f_\pi= 1 + r_2 + \ldots
 \ee
In contrast to the semileptonic decay discussed above,  the Ademollo--Gatto theorem does not apply in this case and 
 $r_2$  is not predicted unambiguously in ChPT. Instead one should fix the low energy constants 
 from, say, the lattice QCD 
 studies of   $  f_K/f_\pi$.  This year many new results with 
either $N_F=2$ and $N_F=2+1$ dynamical quarks and rather light quark masses 
have been presented~\cite{milcfk,davies,rbcfk,fk,jam}. Such  obtained values are summarized 
in fig.~\ref{fk} from which we deduce that the  present  overall accuracy is about $1\%$. 
Note in particular  the new lattice results with 
 $N_F=2+1$ dynamical quarks and pions as light as $280$ MeV~\cite{milcfk,davies}, 
 obtained by using the so-called staggered quarks~\footnote{
 Staggered fermions come in four tastes on the lattice. In the continuum limit  
 the extra degrees of freedom decouple from
 physical predictions.  But, at finite
 lattice spacing, where the data are produced, the taste symmetry is violated 
 and this doublers are removed by hand, namely by taking the fourth root of the staggered
 quark determinant. Theoretically, this procedure has been 
 only confirmed in perturbation theory and is currently a subject 
 of controversies within the lattice QCD community~\cite{kronfeld-creutz}. 
Since the staggered dynamical quarks are computationally cheap, 
 the first results  with $N_F=2+1$
 have been produced by this approach. Thanks to 
  recent progress in algorithm building~\cite{algo},  a safer and hopefully competitive alternatives are possible.}
in which they covered    a broad range of lattice spacings 
(i.e., $a=[0.06, 0.15]$ fm) and  
kept sufficiently large physical volumes (i.e., $m_\pi L\gtrsim 5.0$). 
It should be stressed, however, that the sensitivity of 
$f_K/f_\pi$ to the lighter pions is larger 
than in the computation of $\fp$, and that the chiral extrapolations are 
much more 
demanding in this case~\footnote{In some details,  
effects of chiral logs are not clearly disentangled and
analytic terms (NNLO or NNNLO) are still needed in order to extrapolate from the
simulated sea
quark masses (such as $m_\pi\gtrsim 280$ MeV) to the physical point. 
For example, the two studies of ref.~\cite{milcfk} and of ref.~\cite{davies} with staggered quarks
share the same configurations, but they differ in how to extrapolate to the physical masses.
In the end, this  implies a discrepancy for the central values of $f_K/f_\pi=$ from the two analysis, 
(namely, $f_K/f_\pi=1.197^7_{13}$ and $f_K/f_\pi=1.189(7)$ from ref.~\cite{milcfk} and ref.~\cite{davies}
respectively). In any case, once we symmetrise the error of $f_K/f_\pi$ in~\cite{milcfk}, 
we have $f_K/f_\pi=1.194(10)$ and the two values looks now in good agreement.
On the other hand,
  highly improved staggered fermions
  (HISQ) used in~\cite{davies} for the valence quarks are
  designed to reduce the taste violation effects, which also should
reduce the overall systematic uncertainty. }. Notice also that at Lattice 2007
   preliminary studies with $N_F=2+1$  clover quarks and pion masses 
   $\gtrsim 200$ MeV have been presented 
from either   PACS-CS Collaboration~\cite{jam} and 
ref.~\cite{lel}.  
   With respect to the results obtained with staggered
quarks, the PACS-CS  value of $f_K/f_\pi$ in fig.~\ref{fk} is 
restricted to a single lattice spacing ($a=0.09$ fm) and 
relatively small physical
volume ($m_\pi L\gtrsim 2.9$)~\cite{volume}. For ref.~\cite{lel},
  the final analysis is to be completed.
   
From the present knowledge of $f_K/f_\pi$,  we see in  fig.~\ref{vusuni} 
that  $\vus$ from
$K_{\ell2}$  is in agreement with unitarity too.

\section{Future perspectives}

Here we briefly summarize the probable perspectives of the  $K_{\ell3}$
and $K_{\ell2}$ studies.

The  lepton-universality searches in 
$R_K=\Gamma(K\to\mu\nu)/\Gamma(K\to e\nu))$, could give us some precious  hints  on new physics
scenarios. On one hand, $R_K$ can be predicted to 0.04~\% accuracy in the SM,   while 
on the other hand the Higgs contributions [$\propto (s_R u_L)(e_R \nu^\tau_L)$] ,
arising from lepton violating couplings 
 at large $\tan\beta$, can give an effect  $\sim 1\%$~\cite{paride}. Recent measurements
 of this ratio by NA48~\cite{fantecchi} and KLOE~\cite{spadaro}   reached a percent level of accuracy
which makes it very exciting to see what the NA60~\cite{fantecchi} experiment will 
achieve as they  aim  at lowering the uncertainty to the  per-mil level.

A possible  deviation of 
$\vert V_{us}\vert_{K_{\ell 3}}/\vert V_{us}\vert_{K_{\ell 2}}$ from unity, as
argued in some models of physics beyond SM, represents another exciting avenue 
 for the  future searches~\cite{gino,stern}. Notice that in this case the  hadronic uncertainties
enter through  $(f_K/f_\pi)/f_+(0)$, which will hopefully be reduced by the future lattice 
QCD studies with ever lighter pions, as mentioned in the previous section.
 Moreover,  it would be particularly interesting 
 to compute directly $(f_K/f_\pi)/f_+(0)$ on the lattice, i.e., in the same
set of simulations
%~\footnote{As we discussed here, so far the best estimate of 
%$f_+(0)$  is the one made by RBC-UKQCD~\cite{rbcf0} and   
%on this lattice setup (with pions  $\gtrsim 330$ MeV), they also have 
%a preliminary estimate~\cite{rbcfk} of $f_K/f_\pi$ (see fig.~\ref{fk}). 
%Instead, the
%estimate of $f_K/f_\pi$ with lighter pion masses 
%is from $N_F=2+1$ staggered quarks~\cite{milcfk,davies} 
%but their analysis of $f_+(0)$ is very preliminary~\cite{milcf0}.} 
~\footnote{In particular, since the advanced status of staggered simulations, it would be interesting to improve 
the very preliminary analysis of $f_+(0)$ in~\cite{milcf0}.}.

Before the lattice study of  $(f_K/f_\pi)/f_+(0)$ is made, its value can be obtained by using the Callan-Treiman formula,
\be
 f_0(m^2_K-m^2_\pi)= {f_K/f_\pi \over  f_+(0) } + \,
`` \rm{corrections}\, {\cal{O}}(10^{-3})"\quad,
\ee
and the experimental information on the scalar form factor, $f_0(q^2)$.
This involves an extrapolation of the scalar form factor from the physical region 
 $0 <q^2 <(m_K-m_\pi)^2$ to  $q^2=m^2_K-m^2_\pi$ which can be made by using 
an appropriate dispersion relation, as proposed in ref.~\cite{stern}.
 Experimental groups are now implementing this new  parametrization and the results 
are expected soon~\cite{na48,mario,testa,flaviact}.
 This would not only provide an  important cross-check of lattice QCD estimates,  
 but would also help
 improving the  bound on coupling to the right-handed currents~\cite{stern} and/or scalar
 operators~\cite{gino}.

\section{Conclusions}
Besides a few clean observables~\cite{paride}, future progress 
in leptonic and semileptonic kaon decays will mainly rely on  the improvements in 
lattice QCD. It should be emphasized that over the past few years a tremendous progress in lattice QCD
has already been achieved. Most notably the  
quenched approximation has been removed. However, for the precision 
which we  would like to have, much effort is still needed. It is essential to have 
light sea quarks in order to 
match with the ChPT regime, while keeping the discretisation and finite
volume effects under control and below $1\%$ too.

\section{Acknowledgments}
FlaviaNet is a Marie Curie Research Training Network
supported by the European Union Sixth Framework Programme 
under contract MTRN-CT-2006-035482.

\smallskip

\end{document}